# Fast Patient-Specific Breast CT Dosimetry: 22-Fold Acceleration of Monte Carlo MGD Estimation

Amir Entezam[1,2], Ashkan Pakzad[1], Christopher J. Hall[2], Anton Maksimenko[2], Matthew J. Cameron[2], Adam Round[2], Seyedamir T. Taba[3], Keith A. Nugent[1], Daniel Häusermann[2], Patrick C. Brennan[3], Timur E. Gureyev[1], Harry M. Quiney[1]

[1] School of Physics, The University of Melbourne, Parkville, VIC 3010, Australia

[2] Australian Synchrotron, ANSTO, Clayton, VIC 3168, Australia

[3] Faculty of Medicine and Health Sciences, The University of Sydney, Lidcombe, NSW 2141, Australia

## Abstract

Accurate patient-specific mean glandular dose (MGD) estimation in breast computed tomography (BCT) requires anatomically realistic models, but high-resolution patient-derived phantoms impose high computational demands on Monte Carlo (MC) dosimetry. We previously developed and validated an EGSnrc-based framework for synchrotron propagation-based phase-contrast BCT (PB-PCT) for the first-in-human study at the Australian Synchrotron. With participant imaging commencing, this study aimed to optimize phantom spatial resolution and MC efficiency while maintaining MGD accuracy, for a rapid patient-specific dosimetry. Four patient-specific heterogeneous breast phantoms with varying volume and glandularity were generated from PB-PCT images of mastectomy specimens acquired at 35 keV. The highest-resolution phantom for each specimen served as the reference. Phantoms were downsampled by increasing in-plane voxel size and slice thickness, and MGD was recalculated using EGSnrc/DOSXYZnrc. A 4.5% difference from reference MGD was adopted as the accuracy criterion. Primary photon histories were then reduced to determine the minimum required for acceptable MC precision. A common voxel size of 0.30 x 0.30 x 3.00 mm3 maintained MGD within 4.5% of reference for all four phantoms while reducing voxel number by approximately 20-31-fold. MGD was more sensitive to in-plane voxel size than slice thickness, supporting anisotropic downsampling. Reduced-resolution phantoms enabled a 15-fold reduction in photon histories to 2 x 10^8 while maintaining acceptable statistical uncertainty. Combined optimization reduced computation time from approximately 1600 to 72 min, a 22-fold acceleration. Patient-specific phantoms can be substantially downsampled while maintaining MGD accuracy and markedly reducing computational burden. This optimization provides a foundation for future real-time patient-specific BCT dosimetry.

## 1. Introduction

Breast cancer is the most commonly diagnosed cancer among women worldwide, and imaging plays a central role in its early diagnosis [1-4]. Digital mammography (DM) remains the primary screening modality, while digital breast tomosynthesis (DBT) can improve lesion conspicuity by reducing tissue superposition [5–7]. Nevertheless, both techniques require uncomfortable breast compression and DM is limited by the projection of three-dimensional breast anatomy onto a two-dimensional image [5–9]. These limitations are particularly important in women with dense breasts, where overlapping fibroglandular tissue can obscure lesions and reduce diagnostic sensitivity [3,4]. Dedicated BCT addresses these limitations by providing true three-dimensional imaging with isotropic spatial resolution of the uncompressed breast, eliminating anatomical superposition and enabling accurate volumetric visualization of breast tissue [8-14].

Propagation-based phase-contrast computed tomography (PB-PCT) offers an additional opportunity to improve breast imaging by exploiting X-ray refraction in the breast, in addition to conventional attenuation, to enhance soft-tissue contrast [15,16]. Among phase-contrast approaches, propagation-based imaging is particularly attractive because phase effects are visualized through coherent free-space propagation without requiring complex X-ray optical elements [17,18]. Synchrotron radiation is well suited to this technique because of its high spatial coherence, monochromaticity, high photon flux, and quasi-parallel beam geometry [19]. These properties have supported the translation of synchrotron phase-contrast breast imaging from experimental investigations toward human studies [19,20]. At the IMBL of the Australian Synchrotron, the first-in-human PB-PCT program has progressed toward live-participant imaging, creating a need for accurate and computationally efficient patient-specific radiation dosimetry [20,21].

Radiation exposure in breast imaging is commonly characterized using the MGD, which represents the mean energy absorbed per unit mass of glandular tissue, the breast tissue considered most relevant for radiation-induced carcinogenic risk [22,23]. Because MGD cannot be measured directly *in vivo*, MC radiation transport simulations are widely used for its estimation [24-30]. MGD can be related to the incident air kerma ($K_{air}$) through the normalized glandular dose coefficient, DgN

$$MGD = DgN \times K_{air} \qquad (1)$$

where DgN depends on factors including X-ray energy, breast geometry, tissue composition, skin thickness, and imaging geometry [31,32] and K_air is the incident dose to air that can be measured experimentally. Accurate patient-specific MGD estimation therefore requires not only an accurate computational representation of the imaging system and radiation field, but also an appropriate representation of individual breast anatomy that determines DgN [33,34].

Conventional breast dosimetry has historically largely relied on simplified homogeneous phantoms in which adipose and glandular tissues are represented as a uniform mixture with or without a surrounded skin layer [35-37]. Although these models provide standardized and computationally convenient dose estimates, they do not take into account the heterogeneous spatial distribution of glandular tissue, patient-specific morphology, or local tissue interfaces present in real breasts [38-41]. Glandular tissue is distributed non-uniformly throughout the breast, and previous investigations have demonstrated that assumptions regarding breast composition and tissue distribution can substantially influence calculated MGD [37-42]. Patient-derived heterogeneous breast models have consequently been developed to provide a more anatomically realistic basis for BCT dosimetry. Hernandez et al., for example, demonstrated differences between homogeneous and heterogeneous breast models and showed improved agreement between anatomically realistic heterogeneous models and voxelized patient BCT phantoms [43]. Furthermore, substantial spatial variation in energy deposition has been demonstrated within segmented glandular tissue, emphasizing that glandular location and tissue structure influence dose deposition in addition to the overall amount of glandular tissue [44].

We previously developed and validated a unified EGSnrc-based MC framework for patient-specific breast dosimetry of synchrotron PB-PCT at IMBL, using voxelized breast phantoms generated from PB-PCT images of mastectomy specimens [45-47]. The framework incorporates the specific IMBL beam characteristics, realistic heterogeneous breast anatomy, three-dimensional radiation transport, calculation of MGD, and dosimetric analysis. Application of this framework demonstrated the importance of anatomical heterogeneity, skin thickness, and phantom modelling strategy in MGD estimation [45]. However, with the clinical PB-PCT transition to live participants, patient-specific MC dosimetry must be performed rapidly enough to provide MGD estimates within the clinical imaging workflow, while increasing numbers of patient-specific datasets will also impose greater computational and memory demands. Improving MC computational efficiency is therefore important not only for reducing the computational burden, but more importantly for enabling rapid MGD estimation before diagnostic acquisition, providing a pathway toward real-time patient-specific dosimetry within the PB-PCT program.

A major computational challenge arises in the case of patient-specific breast phantoms with high spatial resolution. Three-dimensional breast datasets may contain millions to tens of millions of voxels, resulting in substantial memory requirements and long MC calculation times. Reducing phantom spatial resolution provides a direct means of decreasing this computational burden by reducing the number of voxels required to represent the breast. More importantly, reducing phantom resolution also reduces the number of primary photon histories required to achieve acceptable MC statistical precision, thereby significantly reducing computation time. However, one needs to balance the spatial resolution of the phantom and the achievable accuracy in MGD estimation. Previous studies have investigated the effects of voxel resolution on dose accuracy and computational efficiency in digital breast tomosynthesis [48], mammography [49], and other MC dosimetry applications, including radiotherapy [50]. In particular, Fedon et al. systematically rebinned patient-specific breast phantoms for DBT and demonstrated that coarser voxel resolutions could substantially reduce MC computation time while maintaining glandular dose within a 4.5% accuracy for most investigated breasts [48].

Increasing voxel dimensions modifies the representation of tissue interfaces and introduces partial-volume effects, potentially altering tissue classification and the spatial distribution of glandular and adipose tissues. Limited reconstruction voxel size has previously been shown to result in loss of glandular-tissue information in tomographic breast images [48]. Despite previous investigations in other breast imaging modalities, the trade-off between phantom spatial resolution, MGD accuracy, and computational efficiency has not been systematically established for patient-specific BCT and, in particular, for synchrotron PB-PCT. BCT differs from DM and DBT through its complete 360-degree rotational acquisition geometry, while the high-resolution volumetric datasets generated by synchrotron PB-PCT introduce substantial computational demands. Importantly, the extent to which in-plane voxel size and through-plane slice thickness can be independently increased while maintaining MGD accuracy remains unclear. Furthermore, it has not been systematically investigated yet whether reduced-resolution breast phantoms allow the number of primary MC photon histories to be substantially reduced while maintaining comparable statistical precision for patient-specific BCT dosimetry. Addressing these questions could provide significant computational benefits beyond those achieved through phantom downsampling alone.

Accordingly, this study systematically investigates the effect of patient-specific phantom spatial resolution on MGD accuracy and MC computational efficiency for synchrotron PB-PCT. The overall objective is to identify a practical reduced-resolution phantom that maintains MGD accuracy within an acceptable criterion of the high-resolution reference while substantially reducing memory requirements and calculation time, thereby establishing a computational basis for rapid and scalable patient-specific dosimetry within the IMBL PB-PCT program and similar use cases at other breast imaging facilities.

## 2. Materials and Methods

### 2.1 Patient-specific breast datasets

Patient-specific digital breast phantoms were generated from high-resolution PB-PCT images of freshly excised human mastectomy specimens acquired at IMBL. The imaging was conducted under a Human Research Ethics protocol approved by Monash Health, and written informed consent was obtained from all donors for imaging of their clinical specimens [51].

Four mastectomy specimens, designated Samples 1-4 (S1-S4), were included in the phantom resolution analysis. The specimens covered a range of breast compositions, with glandularity ranging from approximately 11% to 32%. During the PB-PCT scans, each specimen was positioned without compression within a cylindrical plastic holder with a diameter of approximately 12 cm. The cylinder axis was aligned with the CT rotation axis and perpendicular to the incident X-ray beam.

After the collected PB-PCT scans were processed and reconstructed, the high-resolution reference phantoms were generated with in-plane voxel dimensions set to approximately 0.12 mm and slice thicknesses set to approximately 0.65–0.83 mm, depending on the specimen. Their characteristics are summarized in Table 1.

**Table 1. Characteristics of the high-resolution patient-specific reference breast phantoms.**

| Sample | Reference matrix (voxels) | In-plane voxel size (mm) | Slice thickness (mm) | Reference volume ($cm^3$) | Glandularity (%) |
|---|---|---|---|---|---|
| S1 | 990 × 990 × 80 | 0.12 | 0.68 | 767.77 | 31 |
| S2 | 990 × 990 × 97 | 0.12 | 0.65 | 889.85 | 20 |
| S3 | 990 × 990 × 84 | 0.12 | 0.83 | 983.99 | 10 |
| S4 | 990 × 990 × 80 | 0.12 | 0.83 | 937.13 | 11 |

During PB-PCT acquisition, each specimen was rotated through 360° to obtain a complete tomographic dataset. Imaging was performed using a monochromatic X-ray beam with the energy of 35 keV. Images were acquired using a Teledyne DALSA Xineos 3030HR flat-panel detector with a pixel pitch of 99 μm [51]. Detailed descriptions of the PB-PCT acquisition and image reconstruction procedures have been reported previously [45].

The reconstructed datasets preserved the three-dimensional morphology and heterogeneous internal tissue distribution of the specimens and formed the basis for generating both the high-resolution reference and reduced-resolution patient-specific phantoms.

### 2.3. Generation of patient-specific voxel phantoms

The reconstructed PB-PCT datasets were converted into EGSnrc-compatible voxel phantoms using in-house MATLAB software (R2017b, MathWorks, Natick, MA, USA) together with the CTCREATE utility [52, 53] provided with DOSXYZnrc. In this work, PB-PCT reconstructions produced β-maps that are directly proportional to the local linear attenuation coefficient and describe the spatial distribution of the imaginary component of the complex refractive index within breast tissue. The complex refractive index $n$ is expressed as

$$n=1-\delta+i\beta \quad \text{(Equation 2)}$$

where $\delta$ is the refractive index decrement (real part), responsible for phase shifts, and β is the absorption index (imaginary part), which characterizes X-ray attenuation at a given energy. Consequently, the β value at each voxel reflects the intrinsic absorption properties and composition of the tissue. The β values were converted to HU-equivalent values using Equation (3). The conversion is based on the relation between the β and the linear attenuation coefficient, $\mu = 2k\beta$, where $k$ is the X-ray wavenumber. Substituting this relation into the conventional CT definition of HU yields

$$\mathrm{HU} = 1000\frac{(2k\beta-\mu_{\mathrm{water}})}{\mu_{\mathrm{water}}}. \quad \text{(Equation3)}$$

Where $\mu_{water}$ is the linear attenuation coefficient of water at the corresponding photon energy (28–38 keV), obtained from the NIST database [54].

The HU-based DICOM datasets were saved as coronal slices [45] and processed using CTCREATE to generate voxel-specific material assignments and mass densities. Each voxel was classified as air, adipose tissue, glandular tissue, or skin using the following HU ranges: air, −1000 to −700 HU; adipose tissue, −699 to −100 HU; glandular tissue, −99 to +80 HU; and skin, +81 to +200 HU. The same HU-to-material thresholds were applied to all specimens and investigated spatial resolutions. Air and adipose tissue definitions were obtained from the 521ICRU PEGS4 material library, while glandular tissue and skin were defined using their corresponding elemental compositions and mass densities obtained from NIST [54] and incorporated into the EGSnrc material library using EGS-GUI. The material definitions are summarized in Table 2. Full details of the β-to-HU conversion, CT calibration, phantom-generation procedure, and associated analysis have been reported previously [45] and are therefore only summarized here.

**Table 2. Material definitions used for phantom generation and MC dose calculations.**

| Tissue | EGSnrc material | Density (g cm$^{-3}$) | Elemental composition by mass | Source |
|---|---|---|---|---|
| Air | AIR521ICRU | 0.001204 | N 75.5%, O 23.2%, Ar 1.3% | ICRU 37 |

| Tissue | EGSnrc material | Density (g cm$^{-3}$) | Elemental composition by mass | Source |
|---|---|---|---|---|
| Adipose | ADIPOSE521ICRU | 0.95 | H 11.4%, C 59.8%, N 0.7%, O 27.8%, Na 0.1%, P 0.2% | ICRU 44 |
| Glandular | Glandular | 1.04 | H 10.5%, C 63.0%, N 2.0%, O 23.5%, Na 0.2%, P 0.8% | NIST |
| Skin | Skin | 1.09 | H 10.2%, C 61.5%, N 1.9%, O 26.0%, Na 0.2%, S 0.2% | NIST |

For each specimen, the resulting material and density distributions were written in the EGSnrc egsphant format, containing the three-dimensional voxel boundaries, material assignments, and corresponding mass densities required for DOSXYZnrc dose calculations. The highest-resolution phantom for each specimen was designated as the reference phantom.

**2.4. Phantom resampling and partial-volume assessment**

Starting from each high-resolution dataset, a series of reduced-resolution phantoms was generated by systematically varying the in-plane matrix dimensions and the number of reconstructed slices while preserving the physical dimensions of the breast. Thus, reducing the matrix dimensions increased the physical voxel size without altering the overall external geometry of the phantom.

Through-plane resampling was performed by averaging neighbouring image slices to generate larger effective slice thicknesses, whereas in-plane resampling was performed using bilinear interpolation to obtain the required in-plane voxel dimensions.

Resampling was applied to the quantitative image data before final tissue classification. Consequently, voxels located at tissue interfaces could contain averaged image values from neighbouring tissue types. This approach allowed partial-volume effects and the loss of fine anatomical detail associated with reduced spatial resolution to be incorporated into the resulting phantoms.

Following resampling, the images were classified using the same HU-to-material thresholds as the one applied to the high-resolution reference datasets. The glandularity of each reduced-resolution phantom was checked to be consistent with that of its corresponding reference phantom, thereby minimizing changes in the overall glandular fraction as a confounding factor and allowing the effects of spatial resolution and the associated redistribution of glandular tissue to be evaluated. The resulting material and density distributions were subsequently converted into egsphant files for DOSXYZnrc calculations.

A range of in-plane voxel sizes and slice thicknesses was investigated, both independently and in combination. Initially, one spatial dimension was varied while the other was maintained at or near its reference value, allowing the individual effects of in-plane resolution and slice thickness on MGD to be assessed. As an example, the configurations used for this independent assessment in S3 are presented in Table 3. For the in-plane assessment, the slice thickness was maintained at approximately 0.83–0.84 mm while the in-plane voxel size was progressively increased. Conversely, for the through-plane assessment, the in-plane voxel size was fixed at 0.18 mm while the slice thickness was progressively increased. Based on the observed sensitivity of MGD to each spatial dimension, additional combinations of in-plane voxel size and slice thickness were subsequently selected and investigated.

**Table 3. Configurations used to independently assess the effects of in-plane voxel size and slice thickness for S3.**

| Assessment | Matrix size | In-plane voxel size (mm) | Slice thickness (mm) |
|---|---|---|---|
| In-plane resolution | 990 × 990 × 84 | 0.12 | 0.83 |
| | 635 × 635 × 83 | 0.18 | 0.84 |
| | 400 × 400 × 84 | 0.29 | 0.83 |
| | 200 × 200 × 83 | 0.57 | 0.84 |
| Slice thickness | 635 × 635 × 83 | 0.18 | 0.84 |
| | 635 × 635 × 40 | 0.18 | 1.75 |
| | 635 × 635 × 20 | 0.18 | 3.50 |

The investigated resolutions ranged from values close to the original spatial resolution of the reconstructed PB-PCT dataset to substantially coarser configurations. For each reduced-resolution configuration, the calculated MGD was compared with that obtained using the corresponding high-resolution reference phantom. An MGD difference of approximately 4.5% relative to the reference was chosen as the study-specific criterion for acceptable dosimetric agreement.

### 2.5. MC model of the synchrotron PB-PCT beam

MC simulations were performed using EGSnrc version 10.3. The synchrotron photon field was modelled using BEAMnrc, and three-dimensional energy deposition within the reference and reduced-resolution patient-specific breast phantoms was calculated using DOSXYZnrc. The development and experimental validation of the MC beam model have been described previously [45]; therefore, only parameters directly relevant to the present study are summarized here.

The X-ray beam at IMBL is generated by a superconducting wiggler located approximately 136 m upstream of the experimental imaging position. The experimental beam was expanded and collimated to produce an approximately uniform rectangular field of 20 × 8 $cm^2$, with the breast specimens positioned approximately 2.7 m downstream of the final collimation stage. Rather than explicitly modelling the upstream beamline components, the MC source was defined to reproduce the experimentally delivered photon field at the specimen position, as established previously [45].

In BEAMnrc, the incident radiation field was represented as a spatially uniform, monoenergetic rectangular photon source matching the experimental field dimensions. A photon energy of 35 keV was used throughout this phantom-resolution study. A phase-space file generated at the breast entrance plane was subsequently used as the radiation source for DOSXYZnrc calculations. The phase-space file recorded particle energy, position, direction, and statistical weight and was generated using $5 \times 10^9$ primary photon histories.

Photon transport included photoelectric absorption, coherent (Rayleigh) scattering, incoherent (Compton) scattering, and atomic relaxation. Secondary-electron transport and energy deposition were also included. The electron transport cut-off and production threshold were set to ECUT = AE = 0.521 MeV, corresponding to an electron kinetic energy of 10 keV, while the photon transport cut-off and production threshold were set to PCUT = AP = 0.01 MeV.

All source and transport parameters were maintained consistently throughout the phantom-resolution study. Further details are provided in our previous publication [45].

**2.6. Validation of the MC beam model**

The MC beam model used in this study was previously experimentally validated by comparing measured and simulated percentage-depth-dose distributions over the relevant synchrotron energy range [45,55].

Briefly, experimental depth-dose measurements were obtained using an ionisation chamber positioned within a water-filled PMMA tank, and the corresponding experimental geometry was reproduced in DOSXYZnrc. Agreement between the measured and simulated depth-dose distributions was evaluated using gamma-index analysis, a gold-standard method for comparing dose distributions that simultaneously considers dose difference and spatial agreement, with a 2% dose-difference and 2 mm distance-to-agreement criterion [56,57]. The percentage of evaluated points satisfying this criterion ($\gamma \leq 1$) exceeded 95% across the investigated energies, while MC statistical uncertainties remained below 1% in the high-dose region.

Detailed descriptions of the experimental geometry, detector corrections, water-phantom simulations, and validation procedure have been reported previously [42]. The previously validated beam model was used without modification for the present phantom resolution investigation.

### 2.7. MC BCT dose calculation

Each reference and reduced-resolution patient-specific phantom was imported into DOSXYZnrc for voxel-based energy-deposition scoring. The validated phase-space source was used as the incident radiation source for all breast simulations.

The centre of each phantom was aligned with the simulation isocentre to reproduce experimental PB-PCT geometry. A complete 360° BCT acquisition was represented by rotating the incident radiation field around the phantom in 5° angular increments, corresponding to 72 irradiation angles over one complete rotation.

Energy deposition was scored independently within each phantom voxel and recorded in the standard DOSXYZnrc .3ddose format for subsequent analysis.

During the phantom resolution comparison, the radiation source, photon energy, acquisition geometry, transport parameters, and number of primary photon histories were held constant. Thus, the principal variable between simulations was the spatial representation of the patient-specific phantom.

### 2.8. MGD calculation

For each reference and reduced-resolution phantom, MGD was calculated as the total energy deposited in voxels classified as glandular tissue divided by their total glandular mass:

$$MGD = (\Sigma E_i) / (\Sigma m_i) \quad (2)$$

where $E_i$ is the energy deposited in glandular voxel i and $m_i$ is the corresponding voxel mass. The summations included all voxels classified as glandular tissue. The mass of each glandular voxel was calculated using the assigned glandular-tissue density and the physical volume of voxel i. MGD was calculated for all reference and reduced-resolution configurations. MC statistical uncertainties were obtained from the DOSXYZnrc calculations.

### 2.9. Quantification of phantom-resolution effects

For each specimen, the MGD obtained using the highest-resolution patient-specific phantom was treated as the reference value. The absolute percentage difference between the MGD obtained using each reduced-resolution phantom and its corresponding reference value was calculated.

The effects of phantom resolution were evaluated in relation to in-plane voxel size, slice thickness, and total voxel number. In-plane voxel size and slice thickness were firstly examined separately to determine whether MGD exhibited different sensitivities to resolution changes in the two spatial directions.

The total number of voxels was also evaluated to determine how progressively reducing phantom complexity affected MGD agreement and to identify the extent to which voxel number could be reduced before the MGD difference exceeded the predefined accuracy criterion. Reductions in total voxel number were additionally considered an indicator of reduced phantom size and memory requirements.

An MGD difference of approximately 4.5% relative to the corresponding high-resolution reference was adopted as the primary study-specific criterion for identifying reduced-resolution configurations with acceptable dosimetric agreement. An optimal practical common reduced-resolution phantom configuration was then selected and used to optimize the number of primary photon histories.

### 2.10. Optimization of primary photon histories and computational efficiency

Following selection of the practical reduced-resolution phantom, the number of primary photon histories was progressively reduced to determine the minimum number required to maintain a stable MGD estimate with acceptable MC statistical precision, defined as an MC statistical uncertainty below 0.3%.

The initial simulation used $3 \times 10^9$ primary photon histories, for which MC statistical uncertainties were minimized by the large number of simulated histories. The MGDs for the reduced-resolution phantoms were subsequently simulated using $2 \times 10^9$, $1.5 \times 10^9$, $1 \times 10^9$, $5 \times 10^8$, $2.5 \times 10^8$, $2 \times 10^8$, and $1 \times 10^8$ primary photon histories. All source, geometry, transport, and dose-scoring parameters were otherwise maintained unchanged.

For each history level, MGD and its MC statistical uncertainty were recorded. The minimum acceptable number of primary histories was defined as the lowest tested value for which the MC statistical uncertainty remained below 0.3% and the MGD differed from the $3 \times 10^9$-history reference value by less than 4.5%. History levels below this threshold were considered insufficient when either criterion was exceeded.

The expected statistical behaviour was interpreted using the established MC relationship in which statistical uncertainty is approximately inversely proportional to the square root of the number of independent particle histories:

$$\sigma \propto 1/\sqrt{N} \tag{3}$$

where σ is the MC statistical uncertainty and N is the number of primary photon histories. However, the optimized history number was determined from the observed MGD stability and statistical uncertainty rather than from theoretical scaling alone.

Finally, computational efficiency was subsequently assessed by comparing the wall-clock calculation time of the optimized reduced-resolution simulation with that of the original high-resolution reference using $3 \times 10^9$ histories, while confirming acceptable MC statistical precision. All runtime comparisons were performed in the same computational environment, ensuring that the measured improvement reflected the combined effects of phantom and history-number optimization rather than differences in computing hardware.

## 3. Results

The complete spatial-resolution results for S1–S4 are presented in Table 4, including the reference and downsampled matrix dimensions, total voxel number, in-plane voxel size, slice thickness, MGD, and percentage difference from the corresponding reference MGD. The MGD was normalized to the number of primary photons in the MC simulations. The results were analysed to determine (1) the independent effects of in-plane and through-plane resolution, (2) the effect of combined downsampling, and (3) a practical reduced resolution that balances MGD accuracy and phantom complexity.

**Table4. Effects of phantom spatial resolution on MGD for S1–S4.**

| Sample / Reference | Matrix size | Total voxels (M) | In-plane voxel (mm) | Slice thickness (mm) | MGD (mGy/$10^{15}$ photons) | Difference to reference MGD (%) |
|---|---|---|---|---|---|---|
| **S1 – Reference 990×990×80** | 990×990×80 | 78.41 | 0.12 | 0.68 | 2.31 | 0.00 |
| | 990×990×43 | 42.14 | 0.12 | 1.27 | 2.31 | 0.00 |
| | 635×635×80 | 32.26 | 0.19 | 0.68 | 2.29 | 0.83 |
| | 635×635×54 | 21.77 | 0.19 | 1.01 | 2.27 | 1.73 |
| | 635×635×30 | 12.10 | 0.19 | 1.81 | 2.26 | 2.16 |
| | 450×450×54 | 10.94 | 0.26 | 1.01 | 2.26 | 2.16 |
| | 635×635×20 | 8.06 | 0.19 | 2.72 | 2.25 | 2.60 |
| | 450×450×25 | 5.06 | 0.26 | 2.18 | 2.25 | 2.60 |
| | 500×500×25 | 6.25 | 0.24 | 2.18 | 2.25 | 2.60 |

| Sample / Reference | Matrix size | Total voxels (M) | In-plane voxel (mm) | Slice thickness (mm) | MGD (mGy/$10^{15}$ photons) | Difference to reference MGD (%) |
|---|---|---|---|---|---|---|
| | 400×400×30 | 4.80 | 0.30 | 1.81 | 2.25 | 2.60 |
| | 400×400×20 | 3.20 | 0.30 | 2.72 | 2.22 | 3.89 |
| | 380×380×18 | 2.60 | 0.30 | 3.00 | 2.22 | 3.90 |
| | 200×200×54 | 2.16 | 0.57 | 1.01 | 2.13 | 7.79 |
| | 200×200×20 | 0.80 | 0.57 | 2.72 | 2.13 | 7.79 |
| | 100×100×40 | 0.40 | 1.19 | 1.36 | 2.11 | 8.66 |
| | 100×100×15 | 0.15 | 1.19 | 3.63 | 2.08 | 9.96 |
| | 40×40×10 | 0.02 | 2.97 | 5.44 | 1.82 | 21.21 |
| **S2 – Reference 990×990×97** | 990×990×97 | 95.07 | 0.12 | 0.65 | 2.28 | 0.00 |
| | 990×990×43 | 42.14 | 0.12 | 1.47 | 2.29 | 0.44 |
| | 635×635×97 | 39.11 | 0.18 | 0.65 | 2.29 | 0.44 |
| | 635×635×40 | 16.13 | 0.18 | 1.58 | 2.23 | 2.19 |
| | 635×635×15 | 6.05 | 0.18 | 4.20 | 2.22 | 2.63 |
| | 550×550×43 | 13.01 | 0.21 | 1.47 | 2.20 | 3.51 |
| | 400×400×43 | 6.88 | 0.29 | 1.47 | 2.19 | 3.95 |
| | 400×400×20 | 3.20 | 0.29 | 3.15 | 2.17 | 4.82 |
| | 380×380×21 | 3.03 | 0.30 | 3.00 | 2.18 | 4.39 |
| | 300×300×50 | 4.50 | 0.38 | 1.26 | 2.17 | 4.82 |
| | 300×300×20 | 1.80 | 0.38 | 3.15 | 2.15 | 5.70 |
| | 250×250×43 | 2.69 | 0.46 | 1.47 | 2.14 | 6.14 |
| | 200×200×20 | 0.80 | 0.57 | 3.15 | 2.10 | 7.89 |
| | 100×100×40 | 0.40 | 1.14 | 1.58 | 2.03 | 10.96 |
| | 100×100×15 | 0.15 | 1.14 | 4.20 | 2.00 | 12.28 |
| **S3 – Reference 990×990×84** | 990×990×84 | 82.33 | 0.12 | 0.83 | 2.51 | 0.00 |
| | 990×990×43 | 42.14 | 0.12 | 1.63 | 2.53 | 0.80 |
| | 635×635×83 | 33.47 | 0.18 | 0.84 | 2.49 | 0.80 |
| | 635×635×40 | 16.13 | 0.18 | 1.75 | 2.48 | 1.20 |
| | 635×635×20 | 8.06 | 0.18 | 3.50 | 2.48 | 1.20 |
| | 500×500×40 | 10.00 | 0.23 | 1.75 | 2.45 | 2.39 |
| | 400×400×84 | 13.44 | 0.29 | 0.83 | 2.42 | 3.59 |
| | 430×430×25 | 4.62 | 0.27 | 2.80 | 2.42 | 3.59 |
| | 420×420×23 | 3.32 | 0.30 | 3.00 | 2.42 | 3.59 |
| | 300×300×40 | 3.60 | 0.38 | 1.75 | 2.36 | 5.94 |
| | 300×300×25 | 2.25 | 0.38 | 2.80 | 2.36 | 5.98 |
| | 200×200×83 | 3.32 | 0.57 | 0.84 | 2.29 | 8.62 |
| | 200×200×30 | 1.20 | 0.57 | 2.33 | 2.29 | 8.86 |

| Sample / Reference | Matrix size | Total voxels (M) | In-plane voxel (mm) | Slice thickness (mm) | MGD (mGy/$10^{15}$ photons) | Difference to reference MGD (%) |
|---|---|---|---|---|---|---|
| | 100×100×20 | 0.20 | 1.14 | 3.50 | 2.14 | 14.74 |
| **S4 – Reference 900×900×80** | 900×900×80 | 64.80 | 0.13 | 0.83 | 2.48 | 0.00 |
| | 900×900×43 | 34.83 | 0.13 | 1.53 | 2.47 | 0.47 |
| | 635×635×80 | 32.26 | 0.18 | 0.83 | 2.46 | 0.81 |
| | 900×900×20 | 16.20 | 0.13 | 3.30 | 2.45 | 0.84 |
| | 600×600×60 | 21.60 | 0.19 | 1.10 | 2.44 | 1.61 |
| | 635×635×30 | 10.80 | 0.19 | 2.20 | 2.42 | 2.42 |
| | 635×635×15 | 6.05 | 0.18 | 4.40 | 2.40 | 3.23 |
| | 350×350×66 | 8.09 | 0.33 | 1.00 | 2.37 | 4.44 |
| | 400×400×20 | 3.20 | 0.29 | 3.30 | 2.37 | 4.44 |
| | 380×380×22 | 3.18 | 0.30 | 3.00 | 2.37 | 4.44 |
| | 450×450×15 | 3.04 | 0.25 | 4.40 | 2.36 | 4.84 |
| | 360×360×20 | 2.59 | 0.32 | 3.30 | 2.35 | 5.24 |
| | 300×300×15 | 1.35 | 0.38 | 4.40 | 2.31 | 6.85 |
| | 150×150×20 | 0.45 | 0.76 | 3.30 | 2.23 | 10.08 |
| | 100×100×10 | 0.10 | 1.14 | 6.60 | 2.13 | 14.11 |

### 3.1. Independent effects of in-plane voxel size and slice thickness on MGD

As shown in Table 4, MGD was consistently more sensitive to in-plane downsampling than to increases in slice thickness. For better comparison, Table 5 demonstrates the independent effects of in-plane voxel size and slice thickness on MGD separately.

**Table 5. Independent effects of in-plane voxel size and slice thickness on MG**

| Sample | MGD (mGy/$10^{15}$ photons) | In-plane voxel change | ΔMGD | Slice-thickness change | ΔMGD |
|---|---|---|---|---|---|
| S1 | 2.31 | 0.18 → 0.57 mm (3.2×) | 7.79% | 0.68 → ~2.7 mm (~4.0×) | 2.60% |
| S2 | 2.28 | 0.18 → 0.46 mm (2.6×) | 6.14% | 0.65 → 1.58 mm (2.4×) | 2.19% |
| S3 | 2.51 | 0.18 → 0.57 mm (3.2×) | 8.62% | 0.84 → 3.50 mm (4.2×) | 1.20% |
| S4 | 2.48 | 0.13 → 0.32 mm (2.5×) | 5.24% | ~0.83 → 2.20 mm (~2.7×) | 2.42% |

For example, for S1, increasing the in-plane voxel size from 0.18 to 0.57 mm (3.2-fold), corresponding to the change in matrix sizes of 635×635×54 to 200×200×54, resulted in an MGD difference of 7.79%. In comparison, increasing the slice thickness from 0.68 to 2.70 mm (4.0-fold), corresponding to the matrix zie change 635×635×80 to 635×635×20, resulted in a difference of only 2.60%.

Overall, approximately 2.5–3-fold increases in in-plane voxel size produced MGD differences of 5–9%, whereas comparable or even larger increases in slice thickness generally produced differences of only 1–3%. These results demonstrate that MGD is substantially more sensitive to loss of in-plane resolution than to through-plane downsampling.

### 3.2. Effect of combined phantom downsampling on MGD

Following the independent assessment, combinations of increased in-plane voxel size and slice thickness were investigated across all four patient-specific breast phantoms, as presented in Table 4. Overall, moderate combined downsampling produced relatively small changes in MGD, whereas coarser resolutions, particularly larger in-plane voxel sizes, resulted in progressively greater deviations from the reference MGD.

For S1, configurations with in-plane voxel sizes of approximately 0.23–0.29 mm and slice thicknesses of 1.00–2.16 mm resulted in MGD differences of approximately 2.2–3.9%. At 0.30 × 0.30 × 3.00 $mm^3$, the MGD was 2.22 mGy, corresponding to a 3.90% difference from the reference. Further increases in in-plane voxel size resulted in larger differences, reaching 7.79% at 0.57 mm, 8.66–9.96% at 1.14 mm, and 21.21% with extreme downsampling.

A similar trend was observed for S2. At 0.30 × 0.30 × 3.00 $mm^3$, the MGD was 2.18 mGy, corresponding to a 4.39% difference from the reference value of 2.28 mGy. Coarser in-plane resolutions increased the difference to 6.14% at 0.46 mm, 7.89% at 0.57 mm, and 10.96–12.28% at 1.14 mm.

For S3, several configurations with in-plane voxel sizes up to approximately 0.23-0.27 mm maintained MGD within approximately 4% of the reference. At 0.30 × 0.30 × 3.00 $mm^3$, the MGD was 2.42 mGy, corresponding to a 3.59% difference from the reference value of 2.51 mGy. Increasing the in-plane voxel size to 0.38, 0.57, and 1.14 mm increased the differences to approximately 5.9%, 8.6–8.9%, and 14.74%, respectively.

For S4, moderate downsampling also maintained good agreement with the reference. At 0.30 × 0.30 × 3.00 $mm^3$, the MGD was 2.37 mGy, corresponding to a 4.44% difference from the reference value of 2.48 mGy. Further downsampling increased the difference to 6.85% at 0.38 mm, 10.08% at 0.76 mm, and 14.11% at 1.14 mm.

Based on these results, voxel dimensions 0.30 × 0.30 × 3.00 $mm^3$ were selected as a common practical spatial resolution, as the results were summarized Table 6. At this resolution, the MGD differences were 3.90%, 4.39%, 3.59%, and 4.44% for S1–S4, respectively, with all four phantoms remaining within the study-specific approximately 4.5% criterion. At the same time, the total number of voxels decreased from 78.41, 95.07, 82.33, and 64.80 million in the reference phantoms to 2.60, 3.03, 3.32, and 3.18 million, corresponding to approximately 30-, 31-, 25-, and 20-fold reductions, respectively, with each reduced phantom containing approximately 3 million voxels. This resolution was therefore selected for subsequent MC primary-history optimization.

**Table 6. MGD accuracy and reduction in phantom complexity at the selected common resolution of 0.30 × 0.30 × 3.00 mm$^3$.**

| Sample | MGD (mGy/$10^{15}$ photons) | MGD at 0.30 × 0.30 × 3 mm$^3$ (mGy) | ΔMGD | Reference voxels (M) | Reduced voxels (M) | Voxel-number reduction |
|---|---|---|---|---|---|---|
| S1 | 2.31 | 2.22 | 3.90% | 78.41 | 2.60 | ~30× |
| S2 | 2.28 | 2.18 | 4.39% | 95.07 | 3.03 | ~31× |
| S3 | 2.51 | 2.42 | 3.59% | 82.33 | 3.32 | ~25× |
| S4 | 2.48 | 2.37 | 4.44% | 64.80 | 3.18 | ~20× |

### 3.5. MC history optimization and computational efficiency

Following selection of the 0.30 × 0.30 × 3.00 mm$^3$ phantom resolution, the effect of history reduction was evaluated across all four phantoms (Table 7).

The MC statistical uncertainty remained below 0.2% for all four samples when the number of primary photon histories was reduced from $3×10^9$ to $2×10^8$. Further reduction to $1.5×10^8$ histories increased the uncertainty to approximately 2.1%, while at $1×10^8$ histories the uncertainty exceeded 24%. Therefore, $2×10^8$ primary histories were selected as the practical minimum, representing a 15-fold reduction relative to the original simulations while maintaining MC statistical uncertainty below 0.3%.

The computational benefit was evaluated on a workstation equipped with an Intel® Core™ Ultra 7 165H processor. Using the original high-resolution phantom and primary histories, an MGD calculation required approximately 1600 min. Combining the reduced-resolution phantom with primary histories reduced the computation time to approximately 72 min, corresponding to an overall 22-fold reduction in computation time. Computational improvement exceeded the 15-fold reduction in primary histories, indicating an additional benefit from reducing phantom complexity.

**Table 7. Effect of primary photon history reduction on MC statistical uncertainty**

| Primary histories | Reduction in histories | MC uncertainty (S1–S4) | Computation time | Assessment |
|---|---|---|---|---|
| $3 \times 10^9$ | Reference | <0.1% | ~1600 min | Original reference |
| $1 \times 10^9$ | 3× | <0.1% | ~720 min | Acceptable |
| $5 \times 10^8$ | 6× | <0.2% | ~180 min | Acceptable |
| $2 \times 10^8$ | 15× | <0.3% | ~70 min | **Selected** |
| $1.5 \times 10^8$ | 20× | <2.1% | — | Reduced precision |
| $1 \times 10^8$ | 30× | >24% | — | Unacceptable |

## 4. Discussion

This study demonstrates that the spatial resolution of patient-specific breast phantoms can be substantially reduced for synchrotron PB-PCT MC dosimetry without introducing clinically significant changes in MGD. A common voxel size of $0.30 \times 0.30 \times 3.00$ mm$^3$ maintained MGD within approximately 4.5% of the high-resolution reference across all four investigated mastectomy samples, while reducing the total voxel number by approximately 20 to 31-fold. At the 4 mGy MGD, targeted for participant PB-PCT imaging at IMBL [20,21], a 4.5% difference corresponds to only approximately 0.18 mGy, supporting the practical acceptability of the selected dosimetric accuracy criterion.

The results also showed that MGD was more sensitive to increasing the in-plane voxel size than to increasing the slice thickness, indicating that anisotropic downsampling may be more appropriate than isotropic downsampling for patient-specific breast dosimetry. Increasing voxel dimensions reduces the representation of tissue interfaces and alters the spatial distribution of glandular tissue through partial-volume effects, which can affect energy deposition even when the overall glandular fraction remains similar.

The greater sensitivity to in-plane discretisation may also be related to the PB-PCT irradiation geometry, where photons travel mainly within the transverse planes. Increasing the in-plane voxel size therefore reduces the representation of tissue interfaces along the photon path, whereas increasing slice thickness mainly reduces anatomical resolution in the perpendicular direction.

The benefit of anisotropic downsampling is consistent with the findings of Fedon et al. [45] for patient-specific DBT dosimetry. They similarly demonstrated that anisotropic voxels could preserve glandular-dose accuracy while permitting substantially greater reduction in spatial resolution than isotropic rebinning, and reported that dose sensitivity depended on the direction of discretisation. The present study extends this concept to patient-specific BCT and synchrotron PB-PCT, where the breast is irradiated over a complete 360° acquisition and substantially higher-resolution volumetric datasets must be handled. Despite these differences in imaging geometry, both studies demonstrate that preserving fine resolution equally in all three dimensions is unnecessary for accurate MGD estimation and that appropriately selected anisotropic voxels can provide an effective compromise between anatomical representation and computational efficiency.

The practical significance of this optimization becomes particularly important as the IMBL PB-PCT program transitions to live-participant imaging, where patient-specific MC dosimetry must be performed rapidly enough to provide MGD estimates before the diagnostic acquisition. At the same time, the increasing number of patient-specific datasets will impose greater computational and memory demands. Reducing these requirements is therefore important for integrating patient-specific dosimetry into the clinical imaging workflow and ultimately enabling real-time dose and imaging optimization for the individual patient.

As a future direction, the approximately 22-fold reduction in computation time achieved in this study, together with our previously developed and validated unified EGSnrc-based patient-specific dosimetry framework [45], provides the methodological and computational foundation for progressing toward real-time patient-specific dosimetry. The next step will be to quantitatively investigate the image-quality requirements of a very-low-dose, low-resolution pre-scan needed to generate a reduced-resolution patient-specific phantom of sufficient quality for accurate MGD estimation. This phantom could then be used for rapid MC estimation of MGD before diagnostic acquisition. Implementation on high-performance computing infrastructure could further reduce calculation times toward a clinically practical timescale, potentially allowing the estimated patient-specific MGD to be used to optimize imaging parameters for the individual breast before diagnostic scanning. Importantly, this concept is not limited to synchrotron PB-PCT; the workflow could potentially be adapted to other breast CT systems, providing a transferable approach toward real-time patient-specific dose and imaging optimization.